\documentclass[useAMS,usenatbib,a4paper]{mn2e}
\usepackage{aas_macros}
\usepackage{graphicx}
\usepackage{mathptmx}
\usepackage{amsmath}
\usepackage{amssymb}

\newcommand{\Msun}{\ensuremath{\rmn{M}_\odot}}
\newcommand{\<}{\begin{eqnarray}}
\renewcommand{\>}{\end{eqnarray}} 
\renewcommand{\bar}{\overline}

\renewcommand{\hat}{\widehat}

\renewcommand{\(}{\left(}
\renewcommand{\)}{\right)}

\renewcommand{\d}{\mathrm{d}} 

\title[Stochastic stellar growth]{
On the mass function of stars growing in a flocculent medium
}
\author[Th. Maschberger]
{Th. Maschberger\thanks{e-mail: thomas.maschberger@obs.ujf-grenoble.fr}\\
\small \it 
UJF-Grenoble 1 / CNRS-INSU, Institut de Plan\'etologie et d'Astrophysique de Grenoble (IPAG) UMR 5274, Grenoble, F-38041, France
}
\date{}

\pagerange{\pageref{firstpage}--\pageref{lastpage}} \pubyear{201X}

\begin{document}

\maketitle

\label{firstpage}

\begin{abstract}
Stars form in regions of very inhomogeneous densities and may have chaotic orbital motions.
This leads to a time variation of the accretion rate, which will spread the masses over some mass range.
We investigate the mass distribution functions that arise from fluctuating accretion rates in non-linear accretion, $\dot{m} \propto m^{\alpha}$.
The distribution functions evolve in time and develop a power law tail  attached to a lognormal body, like in numerical simulations of star formation.
Small fluctuations may be modelled by a Gaussian and develop a power-law tail $\propto m^{-\alpha}$ at the high-mass side for $\alpha > 1$ and at the low-mass side for $\alpha < 1$.
Large fluctuations require that their distribution is strictly positive, for example, lognormal.
For positive fluctuations the mass distribution function develops the power-law tail {\it always} at the high-mass hand side, independent of $\alpha$ larger or smaller than unity.

Furthermore, we discuss Bondi--Hoyle accretion in a supersonically turbulent medium, the range of parameters for which non-linear stochastic growth could shape the stellar initial mass function, as well as the effects of a distribution of initial masses and growth times.
\end{abstract}


\section{Introduction}

Star forming regions typically show a very inhomogeneous structure with large  variations in the gas density due to turbulence and filaments.
Thus, accretion rates of forming stars, depending on gas density, will show fluctuations.
Sufficiently substantial fluctuations in the accretion rate will spread out the starting masses after accretion occurred, which affects, like many other effects, the stellar initial mass function, the distribution of stellar masses at their `birth'.
The accretion rate of a point mass in a homogeneous medium follows $\dot{m} \propto \rho m^2$, which is Bondi-Hoyle-Lyttleton accretion \citep{Edgar-2004a}.
In a flocculent medium the density variations change the deterministic Bondi--Hoyle--Lyttleton accretion into a stochastic process, which is multiplicative, as $\dot{m}$ depends on $m$, and non-linear, as $\dot{m} \propto m^2$.
In this paper, we investigate the distribution function that arises from such a non-linear multiplicative stochastic process.

A linear stochastic process, fragmentation, has been employed for some time to explain the shape of the stellar initial mass function at low masses.
The initial mass function has at low masses a lognormal shape, but develops a power-law tail at high masses \citep{Kroupa-2002a,Chabrier-2003b}.
\citet{Larson-1973a}, \citet{ElmegreenMathieu-1983a} and \citet{Zinnecker-1984a} modelled fragmentation as a sequence of discrete fragmentation steps,  during each the mass is reduced by a fraction of itself.
Applying the central limit theorem leads to a lognormal distribution function after a sufficient number of fragmentation steps.
Because the fragment mass is chosen to depend linearly on the mass of the cloud this is a linear process.
Fragmentation alone seems not to embrace the whole star formation process, as the initial mass function has a power-law tail at the massive end.
This deviation from lognormality has been explained, amongst other ideas, by competitive accretion, which is non-linear accretion of the fragments without fluctuations in the accretion rate \citep{Larson-1978a,Zinnecker-1982a,BonnellBateClarke-1997a,BonnellBateClarke-2001a,BonnellClarkeBate-2001a,BateBonnellBromm-2003a}. 
The power-law tail of the initial mass function then arises from the scatter of the initial masses from which accretion starts.

With (linear) fragmentation and (non-linear) accretion two processes encounter each other that have opposite sign.
Fragmentation leads (on average) to a reduction of mass, whereas accretion increases mass.
One can therefore speculate, whether it is possible to model the main part of the star formation process as a single stochastic process that is non-linear.
This has already been attempted by \citet{Marcus-1968a}, who adopted the random splitting model of \citet{Filippov-1961b} (who extended the work of  \citealp{Kolmogorov-1941a}\footnote{In English: \citet{Kolmogorov-1992a}}).
\citet{Marcus-1968a} assumed a time-discrete non-linear stochastic process.
An important aspect of the results by \citet{Marcus-1968a} is that simultaneously the distribution of the number of fragments {\it and} the mass distribution of the fragments are derived.
We investigate a time-continuous non-linear stochastic process, generalizing the work of \citet{Marcus-1968a} in that respect, but do not derive a distribution for the number of fragments.

In linear or non-linear random fragmentation, it is usually assumed that the fraction that is lost during a step follows a Gaussian distribution.
This has a curious side effect: the mass distribution function at the end does not vanish above the initial mass.
But nothing can fragment to a mass larger than its initial mass.
The reason for this lies in the Gaussian distribution used to describe the fragment distribution, there is a non-vanishing probability that some fragments have negative mass.
In stellar growth (which is `negative fragmentation'), the distribution of accretion rates is due to the fluctuations in the gas density.
The gas density cannot be smaller than zero.
Therefore, we investigate the case of a non-linear stochastic process with a strictly positive distribution of fluctuations as well.
The difference to Gaussian noise can be quite substantial.

In order to model the star formation process, it is, besides growth, also necessary to consider the initial mass from which accretion starts and the time how long accretion lasts.
Several models covering all these aspects are present in the literature, for example:
\citet{BasuJones-2005a} assume a lognormal distribution of initial masses with growth $\propto m$ or $\propto m^{2/3}$ and an exponential distribution of growth times.
\citet{BateBonnell-2005a} consider growth with a lognormal distribution of accretion rates without a mass dependence, no distribution of initial masses, and also an exponential distribution of growth times.
Related to this mass-independent growth with lognormal accretion rates is the discussion of stable distributions in
\citet{CartwrightWhitworth-2012a}.
In a series of papers, \citet{Myers-2000a,Myers-2008a,Myers-2009a,Myers-2011a,Myers-2012a}
published increasingly elaborate models of the star formation process.
Their main components are an exponential distribution of growth times, accretion with a mass-independent and a (non-linearly) mass-dependent contribution, and a constant initial mass (a distribution of initial masses is considered in \citealp{Myers-2009a}).
The model of \citet{DibShadmehriPadoan-2010a} contains a distribution of initial masses originating from gravoturbulent accretion (lognormal with a power-law tail) from which stars grow with a mass-dependent accretion rate that is exponentially dampened in time \citep[cf. also][]{DibKimShadmehri-2007a}.
Effects of fluctuating accretion rates are to our knowledge not yet considered in the literature.
In Section \ref{sec_imf}, we describe the effects of a distribution of initial masses and a distribution of growth times on the mass distribution arising from fluctuating accretion rates in a stochastic growth process.

Before embarking on the investigation of non-linear growth with a distribution of the accretion rates we would like to mention that our results may have more applications than stellar growth.
Non-linear stochastic processes appear in many other contexts, particularly in the context of turbulence.
It is, for example, found that the probability distribution function (pdf) of the gas density in a turbulent medium shows power law tails in case of a polytropic index other than unity \citep{PassotVazquez-Semadeni-1998a}.
Given that isothermal turbulence is assumed to be a linear multiplicative process, the occurrence of the power-law tails hints at a non-linear process.
Similarly, the pdf of velocity in a turbulent medium show power-law tails \cite[e.g.][]{KrumholzMcKeeKlein-2006a}, which again may be a sign of non-linearity.
Gravity also leads to non-linear stochastic processes, power-law tails appear in self-gravitating turbulent gas if it is isothermal  \citep[e.g.][]{KlessenBurkert-2000a}.
Even stellar dynamics can be seen as a stochastic process \citep{Chandrasekhar-1943a}.

The outline of this paper is as follows:
In  the next section we prepare the necessary mathematical prerequisites with the help of linear stochastic growth.
Section \ref{section_nonlinear_growth} contains results for non-linear stochastic growth where the random increments are assumed to be Gaussian.
Strictly positive fluctuations are considered in Section \ref{section_positive_noise}.
In Section \ref{section_applications}, we discuss Bondi-Hoyle accretion in a supersonically turbulent medium, the typical parameter ranges that would affect the stellar initial mass function and the effects of a distribution of initial masses and growth times.
The usual summary concludes the paper.

\section{A time-continuous stochastic formulation of linear growth or fragmentation}\label{section_growth}

We start with the simplest form of mass-dependent growth where the accretion rate is linear in mass.
Without fluctuations this is described by the differential equation
\< \frac{\d m }{\d t } &=& A m. \label{lin_ode} \>
The quantity $A$ accounts for all the constants that are involved.
Suppose now that the star grows in a flocculent medium.
Then, the fluctuating gas density will cause fluctuations in the accretion rate, or, more specifically, $A$ will be varying.
The mass dependence remains unchanged.

Mathematically, the fluctuations can be introduced by changing equation \ref{lin_ode} from an ordinary differential equation to a stochastic differential equation,
\< \d m &=& m ( a \d t + b \d W ). \label{lin_sde} \>
Now $A$ is split into two terms, $a \d t$, which describes the mean of $A$, and $b \d W$, which describes the fluctuations around the mean.
We assume that the fluctuations stem from a normal (or Gaussian) distribution,
\< \mathcal{N} (x;\mu,\sigma) &=& \frac{1}{\sqrt{2 \upi} \sigma} e^{ - \frac{1}{2} \frac{ ( x - \mu)^2}{\sigma^2} }, \>
with zero mean and variance $\d t$, so that the distribution of $\d W$ is 
\< p (\d W) &=& \mathcal{N} (\d W; \mu=0, \sigma = \sqrt{ \d t }).\>
For $\d t=1$, the distribution of $\d W$ has unit variance.
Multiplying $b$ and $\d W$ is equivalent to scaling the variance to the desired amount of fluctuations.
Alternatively, the distribution of accretion rates can be written as 
\< p (A) &=& \mathcal{N} (A ; \mu = a  , \sigma = b ). \>

For integrating equation \ref{lin_sde} we need to establish how the integral $\int_0^t \d W$ is to be interpreted.
The integral can be approximated with the limit  of $\sum_{i=1}^n \d W_i$, where all the $\d W_i$ are independently drawn from a normal distribution.
The normal distribution has the property that the sum of two normal variates follows again a normal distribution (in other words, the normal distribution is infinitely divisible).
The parameters for the sum variate are $\mu = \mu_1 + \mu_2$ and $\sigma^2 = \sigma_1^2 + \sigma_2^2$.
Therefore $\sum_{i=1}^n \d W_i$ will again obey a normal distribution with $\sigma^2 = \sum_{i=1}^n \sigma_i^2 = \sum_{i=1}^n \( \sqrt{\d t} \)^2 = t $. 
The limit $\d t \to 0$ and $n \to \infty$ gives the integral $W_t= \int_0^t \d W$, which is a normally distributed random number with zero mean and variance $t$.
A more rigorous derivation this can be found in the literature on stochastic differential equations \citep[e.g.][]{Oksendal-2002a}.

The stochastic calculus is not unique, there are two ways of defining it, the It\=o and the Stratonovich calculus.
Depending on the nature of the fluctuations, the one or the other are more appropriate.
In our case, the fluctuations in density are not caused by the growth process, but by external effects (e.g. turbulence or a chaotic motion of the growing star).
Such external fluctuations require the Stratonovich calculus \citep{Kampen-2006a}, which preserves the standard rules of calculus.

With this we can solve equation \ref{lin_sde} by integrating it to
\< \log m - \log m_0 &=& a t + b W_t,  \label{lin_sde_solution} \>
where $m_0$ is the initial mass and growth starts at $t=0$.

For a population of growing stars, the distribution of $m(t)$ can be found by noting that 
\< a t + b W_t &\sim& \mathcal{N} \(\mu = a t, \sigma = b \sqrt{t} \) \>
($\sim$ denotes `is distributed as').
Thus,
\< \log m - \log m_0 &\sim& \mathcal{N} \(\mu = a t, \sigma = b \sqrt{t} \), \>
or,
\< p (m) &=& \frac{1}{m} \frac{1}{\sqrt{2 \upi}} \frac{1}{b \sqrt{t} } e^{ - \frac{1}{2} \frac{ \( \log m - \log m_0 - a t \)^2}{ b^2 t } }, \label{lin_sde_pdf} \>
where the factor $1/m$ comes from the transformation from $\log m$ to $m$.
The distribution of $m(t)$ is, like in the discrete case, a lognormal distribution.

The distribution function of $m(t)$ (equation \ref{lin_sde_pdf}) is also defined for values  $<m_0$, as in the discrete case. 
$\d m$ can have values $< -m$, as $\d W$ can reach up to $-$ infinity.
Thus,  the mathematical formulation of the growth/fragmentation process, both continuous and discrete, is not strictly correct.
Nevertheless, for a  small level of fluctuations the probability for $m < m_0$ can be very small.
Note that $m(t)$ is always larger than zero.

Another puzzling property of equation \ref{lin_sde_pdf} lies in the expectation value of $m(t)$,
\< E(m_t) &=& 
m_0 e^{ a t  + \frac{b^2}{2} t }.
\>
Here an additional term $\frac{b^2}{2} t$ appears compared to the solution 
$ m (t) = m_0 e^{ A t} $
of the deterministic differential equation \ref{lin_ode}.
This `spurious drift' is a consequence of using the Stratonovich stochastic calculus.

\section{Non-linear growth}\label{section_nonlinear_growth}

\begin{figure}
\begin{center}
\includegraphics[width=8cm]{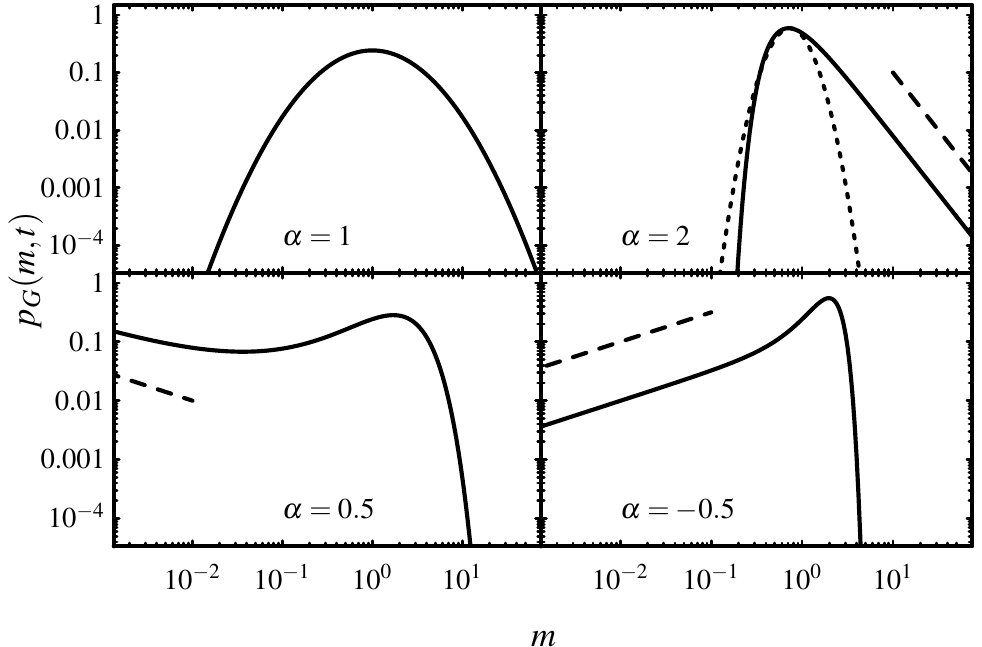}
\end{center}
\caption{\label{figure_nonlin_gauss_solutions}
Comparison of the distribution functions $p_\text{G} (m,t)$ for various $\alpha$.
The parameters are $a=1$, $b=1$, $t=1$, $t_0=0$ and $m_0 = 1$.
For $\alpha = 1$ (top left) the solution is a lognormal distribution.
In the case of $\alpha > 1$ (top right), a power-law tail at high masses develops (dashed).
The dotted line is a lognormal distribution with parameters chosen to fit the lower part of $p_\text{G}$.
If $ 0 < \alpha < 1$ (bottom left) a falling power-law tail at low masses develops, whereas if $\alpha  < 0$ the power-law increases.
}
\end{figure}

The accretion of stars is not linear in mass.
Depending on the environment  $\dot{m} \propto m^2$ (Bondi-Hoyle accretion)  or  $\dot{m} \propto m^{2/3}$ (gas-dominated potential, \citealp{BonnellClarkeBate-2001a}).
Small fluctuations in the accretion rate due to a flocculent density or random stellar motions can be described  neglecting that $\d W $ can take negative values, the corresponding stochastic differential equation is
\< \d m &=& m^{\alpha} ( a\ \d t + b  \d W ).  \label{nonlin_sde} \>
This can be solved proceeding analogously to the previous section.
In this section we discuss the solution of equation \ref{nonlin_sde} and its properties.
For an application to stellar accretion with large fluctuations we need modify equation \ref{nonlin_sde}, see the next section.

The solution for $m(t)$ is 
\< m(t) &=& \( (1-\alpha) \( \frac{m_0^{1-\alpha}}{1-\alpha} + a t + b W_t \) \)^{\frac{1}{1-\alpha}}, \label{nonlin_sde_solution_m} \>
which reduces for $b=0$ to the solution of the deterministic growth law.

For exponents $\alpha > 1$, we have to account for the fact that the solutions are exploding in the deterministic case ($b=0$), infinite masses are reached within a finite time,
\< t_\text{ex} &=& \frac{m_0^{1-\alpha}}{a (\alpha -1) }. \>
When fluctuations are added to the accretion rate, then there is no single explosion time any more.
Depending on the particular fluctuations that are encountered by a growing star it may explode earlier or later, at some random time.
$m(t)$ of equation \ref{nonlin_sde_solution_m} becomes undefined for large $W_t$.
Thus, we have to require 
\< W_t < - \frac{1}{b} \(  \frac{m_0^{1-\alpha}}{1-\alpha} + a t  \) =: u_t \>
if we consider only the not yet exploded particles.
At some time $t$ the fraction of the population that is not yet exploded is 
\< f_\text{nex, G} (t) = \Phi (u_t; 0, \sqrt{t} ), \>
($\Phi$ is the cumulative normal distribution), because $W_t \sim \mathcal{N} (0, \sqrt{t})$.
In nature the exploding solutions are suppressed because an infinite reservoir from which material could be accreted does not exist.
This should be accounted for in the growth model, for example in the line of logistic growth.
Massive stars (which would become an exploding solution) exercise a strong feedback as well, which likewise suppresses exceedingly large accretion rates.
Unfortunately, accounting for a finite reservoir and feedback is beyond the scope of this paper.

The mass distribution function for the population is then
\< p_\text{G} (m, t) &=&
\frac{1}{f_\text{nex, G} (t) }
 \frac{1}{m^\alpha} \frac{1}{\sqrt{2 \upi}} \frac{1}{b \sqrt{t} } 
e^{ - \frac{1}{2} \frac{ \(  \frac{1}{1-\alpha} \( m^{1-\alpha} -m_0^{1-\alpha} \) - a t \)^2 }{ b^2 t } }.
\label{nonlin_sde_pdf}
\>
The factor $1/ f_\text{nex, G} (t) $ normalizes $p_\text{G}$ as probability by accounting for the exploded solutions if $\alpha > 1$.
For $\alpha < 1$, it is not required, $f_\text{nex, G} (t)  = 1$.
Fig. \ref{figure_nonlin_gauss_solutions} shows $p_\text{G}$ for various values of $\alpha$. 
The other parameters are $a=1$, $b=1$, $t=1$, and $m_0=1$. 
For $\alpha=1$ (top-left panel), the solution corresponds to a lognormal distribution, as discussed in Section \ref{section_growth}.
For $\alpha \neq 1$ (and $\alpha \neq 0$), a power-law tail appears at one side of $p_\text{G}$.
If $\alpha > 1$ (top-right panel), the power-law tail $\propto m^{-\alpha}$ develops at the high-mass end of $p_\text{G}$.
The power law  is indicated by the dashed line.
At smaller masses, $p_\text{G}$ behaves similar to a lognormal distribution which is shown by the dotted curve for comparison (the parameters for the lognormal distribution are chosen to follow $p_\text{G}$).
For exponents smaller than unity, in the growth law a power law tail develops at the left hand side of the peak at small masses.
The other side of the distribution resembles a lognormal distribution.
If $0 < \alpha < 1$, the power law is decreasing, which gives the distribution a rather peculiar appearance.
For $\alpha < 0$, i.e. for example $\dot{m} \propto 1/m$, the low-mass power law increases, so that $p_\text{G}$ appears mirrored to the case of $\alpha > 1$.
Although $\alpha < 0$ may not be relevant for stellar accretion, it might be suitable to describe the stochastic process for the distribution of velocities in supersonic turbulence.
\citet{KrumholzMcKeeKlein-2006a} found in numerical simulations that the distribution of velocities has a power-law tail at the left hand side.

We do not discuss at this point the time evolution and the behaviour of the mean mass, but postpone it to the next section for a comparison with the case of strictly positive fluctuations.

\section{Strictly positive noise}\label{section_positive_noise}

\begin{figure}
\begin{center}
\includegraphics[width=6cm]{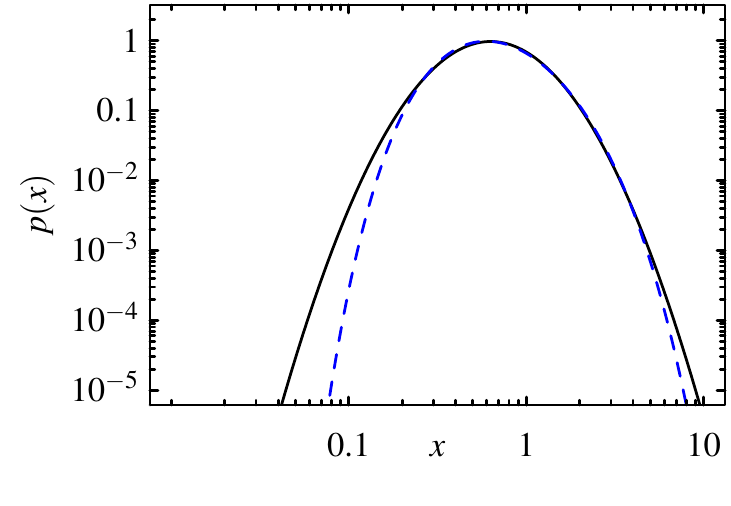}
\end{center}
\caption{ \label{figure_invgauss_lognormal}
Comparison of a lognormal distribution (blue dashed curve) and an inverse Gaussian distribution (black solid) as approximation.
Both have the same mean ($1$) and variance (0.6).
The lognormal distribution is over a wide range well approximated by the inverse Gaussian distribution.
}
\end{figure}

The fluctuations in the stellar accretion rates originate in the variations of the gas density.
If there are only small variations, then their description with a mean density modulated by a Gaussian is sufficiently accurate.
However, if the gas density is very flocculent, as for example in supersonic turbulence, then the use of a Gaussian introduces an undesirable side-effect: the random variate describing the fluctuations can become so large that it exceeds the mean density.
A strongly under-dense region would then be assigned a negative density, which is physically impossible.
In supersonic turbulence, the gas density pdf has been found to follow a lognormal distribution,
\< p_{l \mathcal{N} }  (x) &=& \frac{1}{x} \frac{1}{\sqrt{2 \upi } \sigma_l} e^{- \frac{1}{2} \frac{\( \log x - \mu_l\)^2}{\sigma_l^2} } \label{lognormal_pdf} \>
\citep[e.g. ][]{Vazquez-Semadeni-1994a,PassotVazquez-Semadeni-1998a,NordlundPadoan-1999a}.
We could use the lognormal distribution to describe the fluctuations.
The lognormal distribution is, like the normal distribution, infinitely divisible \citep{Thorin-1977a} and consequently is suitable to describe fluctuations in a stochastic differential equation.
However, there are no formulas for the sum distribution of two lognormal variates, so the stochastic integral cannot be solved analytically.

Thus, we use for practical reasons an approximation of the lognormal distribution, the inverse Gaussian distribution (invGauss),
\< p_\text{invGauss} (x; \nu,\lambda) &=& \( \frac{\lambda}{2 \upi x^3} \)^{\frac{1}{2}} e^{- \frac{\lambda (x-\nu)^2 }{2 \nu^2 x } }. \>
It has expectation value  $E(x) = \nu$ and  variance  $\text{Var} (x) = \nu^3/\lambda$.
A comparison of the invGauss and the lognormal distribution with the same expectation value and variance is shown in Fig. \ref{figure_invgauss_lognormal}. 
Compared to the invGauss the lognormal has somewhat heavier tails.

\begin{figure}
\begin{center}
\includegraphics[width=8cm]{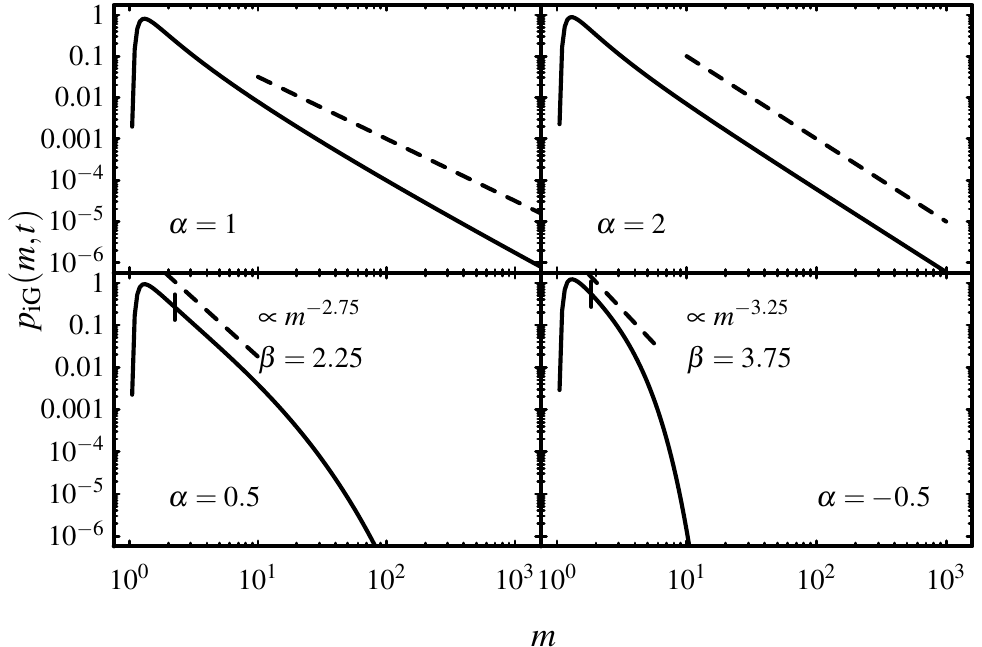}
\end{center}
\caption{\label{figure_nonlin_invgauss_solutions}
Distribution functions $p_\text{iG} (m,t)$ for various $\alpha$ (compare with Fig. \ref{figure_nonlin_gauss_solutions}).
The parameters are $a=1$, $b=1$, $t=1$, $t_0=0$ and $m_0 = 1$.
$p_\text{iG}$ has a power-law tail for $\alpha \geq 1$ and behaves even for $\alpha < 1$  similar to  a power law over some mass range.
The dashed lines show power laws whose exponents are discussed in the text.
}
\end{figure}

As model of strictly positive fluctuations, we choose that an infinitesimal fluctuation 
\< \d \text{iG}_{a,b} & \sim & \text{invGauss} \( \nu = a \d t , \lambda = \frac{a^3}{b^2} (\d t)^2 \). \>
It has the mean value $a \d t$ and variance $ b^2 \d t$, like fluctuations from a normal distribution.
For the approximation of the integral by a sum we note that the sum of two invGauss random numbers with the same $\nu$ and $\lambda$
follows again an invGauss distribution with $\nu' = 2 \nu$ and $\lambda' = 2^2 \lambda$.
With this follows that
\< \text{iG}_{a,b,t} =:  \sum_{i=1}^n \d \text{iG}_{a,b} &\!\!\!\!\!\!\! \sim \!\!\!\!\!\!& \text{invGauss} \( \nu = a n \d t , \lambda = \frac{a^3}{b^2} ( n \d t )^2 \), \>
the sum of infinitesimal fluctuations follows again an invGauss distribution.
The mean of $\text{iG}_{a,b,t}$ is $a n \d t = a t$ and the variance is $b^2 (n \d t)^2 = b^2 t$.
This also corresponds to the results from the case with a normal distribution.
Now we are able to perform the limits $\d t \to 0$ and $n \to \infty$ which gives the integral $\int \d \text{iG}_{a,b}$.

With this we can pose the mass-dependent growth equation with strictly positive fluctuations,
\< \d m &=& m^\alpha \d \text{iG}_{a,b }.  \>
Using standard calculus, we find for $\alpha \neq 1$ that 
\< m(t) &=& \( (1-\alpha)  \( \frac{m_0^{1-\alpha}}{1-\alpha} +  \text{iG}_{a,b,t} \)  \)^{\frac{1}{1-\alpha}}.  
\>
(If $\alpha = 1$, the term $ m_0^{1-\alpha}/(1-\alpha)$ has to be replaced by $\log m_0$).
Again, we have to consider the exploded solutions if $\alpha > 1$ and need to require
$ \frac{m_0^{1-\alpha}}{1-\alpha} +   \text{iG}_ {a,b,t} < 0$ so that $m(t)$ is not infinite.
The not exploding fraction is then
\< f_\text{nex\ iG} &=& 
P_\text{invGauss} \( - \frac{m_0^{1-\alpha}}{1-\alpha}, \nu=a t,\lambda = \frac{a^3}{b^2} t^2 \)
 \>
where $P_\text{invGauss}$ is the invGauss cumulative distribution function.
For $\alpha \leq 1$ nothing explodes and $f_\text{nex\ iG} = 1$.
With $m(t)$ and $f_\text{nex, iG}$ we can write the mass function,
\< p_\text{iG} (m,t) &\!\!\!\!\! = \!\!\!\!\!&
\frac{1}{f_\text{nex, iG}} 
\frac{1}{m^\alpha}
\frac{1}{\sqrt{2\upi}}
\frac{t}{b} 
\( \frac{X}{a} \)^{- \frac{3}{2}} 
e^{  \displaystyle - \frac{1}{2} \frac{a}{b^2} \frac{ ( X-at)^2}{X} },
\>
where 
\< X &:=& \begin{cases} 
\displaystyle  \frac{m^{1-\alpha}}{1-\alpha} - \frac{m_0^{1-\alpha}}{1-\alpha}  & \alpha > 1 \\
 \log m - \log m_0 & \alpha=1
 \end{cases}. \>

\begin{figure}
\includegraphics[width=8cm]{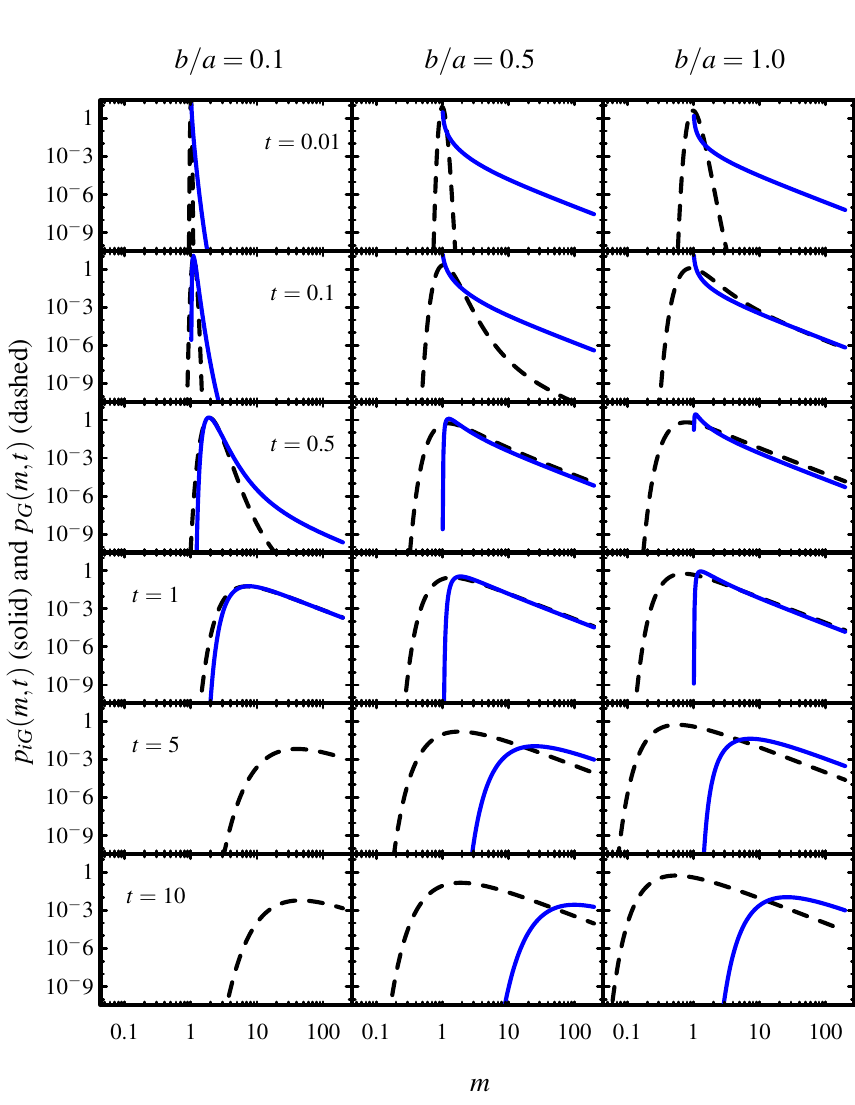}
\caption{ \label{figure_nonlin_invgauss_pdf_time}
Time evolution of the mass functions $p_\text{iG}$ (solid curve) and  $p_\text{G}$  (dashed curve) for $\alpha=2$, $a=1$, $m_0=1$ and $t_0=0$.
In contrast to the Gaussian noise there is  no `growth' below $m_0=1$ for strictly positive noise.
For $b/a=0.1$, $p_\text{iG}$ lies outside the plotting area at $t=5$ and $t=10$.
}
\end{figure}

Fig. \ref{figure_nonlin_invgauss_solutions} shows the behaviour of $p_\text{iG}$ for different $\alpha$ with $a=1$, $b=1$, $m_0=1$, $t=1$ and $t_0=0$.
These are the same parameters as in Fig. \ref{figure_nonlin_gauss_solutions}, which shows $p_\text{G}$.
For $\alpha > 1$, the positive fluctuations lead to the same power law $\propto m^{-\alpha}$ as Gaussian fluctuations.
If $\alpha \leq 1$, the positive fluctuations change the behaviour compared to Gaussian fluctuations, now appears power-law-like behaviour at high masses as well.
This can be characterised with the `exponent' function ($p_\text{iG} \propto m^{S(m)}$),
\< S (m) &=& \frac{\d \log p_\text{iG} }{\d \log m} \\
&=& - \alpha + m^{1-\alpha} \( - \frac{1}{2} \frac{a}{b^2} - \frac{3}{2} \frac{1}{X} + \frac{1}{2} \frac{a^3}{b^2} t^2 \frac{1}{X^2} \).
\> 
$\alpha=1$ leads to a strict power-law behaviour at high masses with exponent $-\alpha - \frac{1}{2}\frac{a}{b^2}$.
However, as visible in Fig. \ref{figure_nonlin_invgauss_solutions} this exponent occurs only as a limiting case and can set in not until very large masses.
At smaller masses $p_\text{iG}$ is steeper.

If $\alpha < 1$, then $S(m)$ has $-\infty$ as limit for $m \to \infty$, $p_\text{iG}$ decays at high masses.
Nevertheless, $p_\text{iG}$ can for some combinations of $a$, $b$ and $t$ be well described by a power law over some mass range above the peak .
The terms $-\frac{1}{2} \frac{a}{b^2} m^{1-\alpha} $ and $\frac{1}{2} \frac{a^3}{b^2} \frac{m^{1-\alpha}}{X^2}$ cancel each other around
\< m_\beta &=&  \(  (1-\alpha) a t  + m_0^{1-\alpha} \)^{\frac{1}{1-\alpha} }, \>
where 
\< p_\text{iG} & \propto & m^{- \alpha - \beta} \>
with
\< \beta &=& \frac{3}{2} \frac{1}{1 - \frac{m_0}{m_\beta}^{1-\alpha} }. \>

Fig. \ref{figure_nonlin_invgauss_pdf_time} displays the time-evolution of $p_\text{iG}$ (solid), as well as $p_\text{G}$ (dashed) for $a=1$ and the ratios $b/a=0.1$, $b/a=0.5$ and $b/a=1.0$.
Both distribution develop a power law tail over time, which happens fastest if the amount of fluctuations is not too large (i.e. $b/a=0.5$).
If the fluctuations are very large, then large accretion rates are occurring only very rarely, as they do if the fluctuations are very small.
Thus, the power-law tail is slower populated.
Both distributions also shift in mass range over time, which is slower for larger $b/a$.
As $p_\text{iG}$ does not allow for `negative' growth it always moves to higher masses.
The Gaussian fluctuations in $p_\text{G}$ make the mass shift so slow for $b/a=0.5$ that the peak effectively does not move.
For $b/a=1$, the shift has reversed, the peak of $p_\text{G}$ moves towards masses smaller than $m_0$.

\begin{figure}
\begin{center}
\includegraphics[width=6cm]{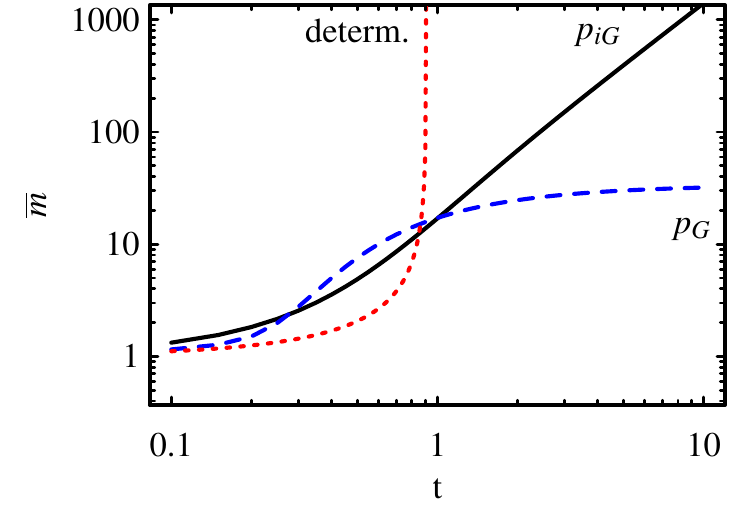}
\end{center}
\caption{ \label{figure_nonlin_invgauss_mean}
Time evolution of the mean of $p_\text{iG}$ (solid curve).
The blue dashed curve shows the expectation for $p_\text{G}$ and the red dotted line the expectation of the deterministic case.
}
\end{figure}

The time development of the expectation value of $p_\text{iG}$ and $p_\text{G}$ (calculated numerically) is given in Fig. \ref{figure_nonlin_invgauss_mean}.
The parameters are $\alpha=2.1$, $a=1$, $b=0.5$, and $m_0=1$.
Also displayed is the deterministic solution as dotted curve which explodes at $t_\text{ex}=0.909$. 
Note that for $1 < \alpha \le 2$ the mean of $p_\text{G}$ and $p_\text{iG}$ is infinity, because the mean for a power law with such an exponent is infinite.
For $t< t_\text{ex}$, the mean mass with a fluctuating accretion rate is larger than expected from deterministic growth.
This is similar to the spurious drift introduced by the stochastic formulation discussed in Section \ref{section_growth}.
Solutions can explode but are not accounted for in the distribution function, so that both averages are finite even for $t > t_\text{ex}$.
The mean $p_\text{iG}$ appears to explode, in contrast to $p_\text{G}$, albeit slower than deterministic.

\section{Application to star formation}\label{section_applications}

\subsection{Bondi-Hoyle accretion in a medium with  supersonic turbulence}

\begin{table}
\caption{\label{table_bondi_hoyle}
Parameters for the distribution of accretion rates of Bondi--Hoyle accretion in a turbulent medium (calculated from the results of \citealp{KrumholzMcKeeKlein-2006a}).
The first two columns give Mach number $\mathcal{M} $ and the ratio of Bondi Radius $r_\text{B}$ to the extent of the region $l$.
The next three columns are for the dimensionless case where  $\rho=1$, $G=1$, $c_\text{s}=1$ and $m = 1$.
The other columns are for physical units where $c_s=220\ \text{m}\ \text{s}^{-1}$.
}
\setlength{\tabcolsep}{.8ex}
\begin{tabular}{lr|ccc|ccc|cc}
     &             &      \multicolumn{3}{c}{  $\rho=1$
} & \multicolumn{3}{c}{  $\rho=1000\ \Msun \text{pc}^{-3}$
}  & \multicolumn{2}{c}{  $\rho =10^4\ \Msun \text{pc}^{-3}$ } \\ 
$\mathcal{M}$ & $\log r_\text{B}/l$ & $a_0$ & $b_0$ & $b_0/a_0$ & 
$a$ &  $b$ &  $b/a$ & $a$ &  $b/a$ \\
 & & \multicolumn{3}{c}{dimensionless} & \multicolumn{5}{c}{$(\Msun/\text{Myr})$} \\
3   &  -5.00  & 0.17 & 1.8 & 11 & 1.8 & 6 & 3.4 & 18 & 1.1 \\ 
3   &  -3.00  & 0.17 & 1.8 & 11 & 1.8 & 6 & 3.4 & 18 & 1.1 \\ 
3   &  -1.00  & 0.071 & 0.42 & 6 & 0.75 & 1.3 & 1.8 & 7.5 & 0.57 \\ 
3   &  1.00  & 0.0029 & 0.014 & 5 & 0.031 & 0.045 & 1.5 & 0.31 & 0.46 \\ 
5   &  -5.00  & 0.07 & 5.4 & 77 & 0.73 & 17 & 24 & 7.3 & 7.5 \\ 
5   &  -3.00  & 0.066 & 4.9 & 74 & 0.7 & 16 & 23 & 7 & 7.2 \\ 
5   &  -1.00  & 0.028 & 0.93 & 33 & 0.3 & 3 & 10 & 3 & 3.2 \\ 
5   &  1.00  & 0.0017 & 0.024 & 15 & 0.017 & 0.078 & 4.5 & 0.17 & 1.4 \\ 
10   &  -5.00  & 0.016 & 10 & 619 & 0.17 & 33 & 191 & 1.7 & 60 \\ 
10   &  -3.00  & 0.016 & 9.1 & 588 & 0.16 & 30 & 181  & 1.6 & 57 \\ 
10   &  -1.00  & 0.0077 & 2.3 & 300 & 0.081 & 7.5 & 93 & 0.81 & 29 \\ 
10   &  1.00  & 0.00083 & 0.072 & 87 & 0.0087 & 0.23 & 23 & 0.087 & 8.5 \\ \end{tabular}
\end{table}

\citet{KrumholzMcKeeKlein-2006a} have studied the distribution of accretion rates for Bondi-Hoyle accretion in a supersonically turbulent medium.
They performed hydrodynamical calculations of accretion on to stationary point masses in a medium that is not self-gravitating.
They found that the normalized accretion rate, $\dot{m}/\dot{m_0}$, can be fitted by a lognormal distribution with parameters depending on the Mach number and the ratio of the Bondi radius, $r_\text{B}$, to the extent of the region, $l$.
The normalization constant is 
\< \dot{m_0} &=& 4 \upi \rho \frac{G^2}{\( \mathcal{M} c_s \)^3 } m^2, \>
with the gas density $\rho$, Mach number $\mathcal{M}$, sound speed $c_s$, and the gravitational constant $G$.
\citet{KrumholzMcKeeKlein-2006a} give their results in dimensionless units where $\rho=1$, $G=1$ and $c_s=1$, and used $m=\frac{13}{32}$.
In Table \ref{table_bondi_hoyle} we give their results adapted to our formalism, where the average accretion rate $a$ and its standard deviation $b$ do not depend on mass.
Typically the ratio $b/a$ in dimensionless units is much larger than unity and can reach values of a few hundred.

Table \ref{table_bondi_hoyle} gives also $a$ and $b$ in physical units (solar masses and million years) for a typical star forming region.
We assume a temperature of 10 K, which corresponds to a sound speed of 220 m/s.
The Bondi radius at that temperature is $r_B=G m c_s^{-2} = 0.09 \ m/\Msun\ \text{pc} $, for a typical region size of 1 pc and 1 \Msun\ mass $\log (r_B/l) = -2.4$.
Although $b/a$ becomes smaller with the scaling it is still typically larger than unity.
This large level of fluctuations will only spread out the initial masses over a range of final masses if accretion lasts for several million years, or the initial masses are $\gg 1\ \Msun$, or the average gas density is very high.

The large fluctuations in the accretion rate do not change the result of \citet{KrumholzMcKeeKlein-2005b} finding that Bondi-Hoyle accretion should not generate large mass increases in typical star forming regions.
For a low average accretion rate strong accretion events are too rare to have an effect.
However, \cite{BonnellBate-2006b} argue that in the central regions of star forming regions the average gas densities are several orders of magnitude higher than the average gas density used by \citet{KrumholzMcKeeKlein-2005b}, in which case accretion does lead to mass growth and, as shown above, a spreading out of the initial masses.

\subsection{The required amount of fluctuations for an impact of non-linear stochastic growth}

\begin{figure*}
\includegraphics[width=8.5cm]{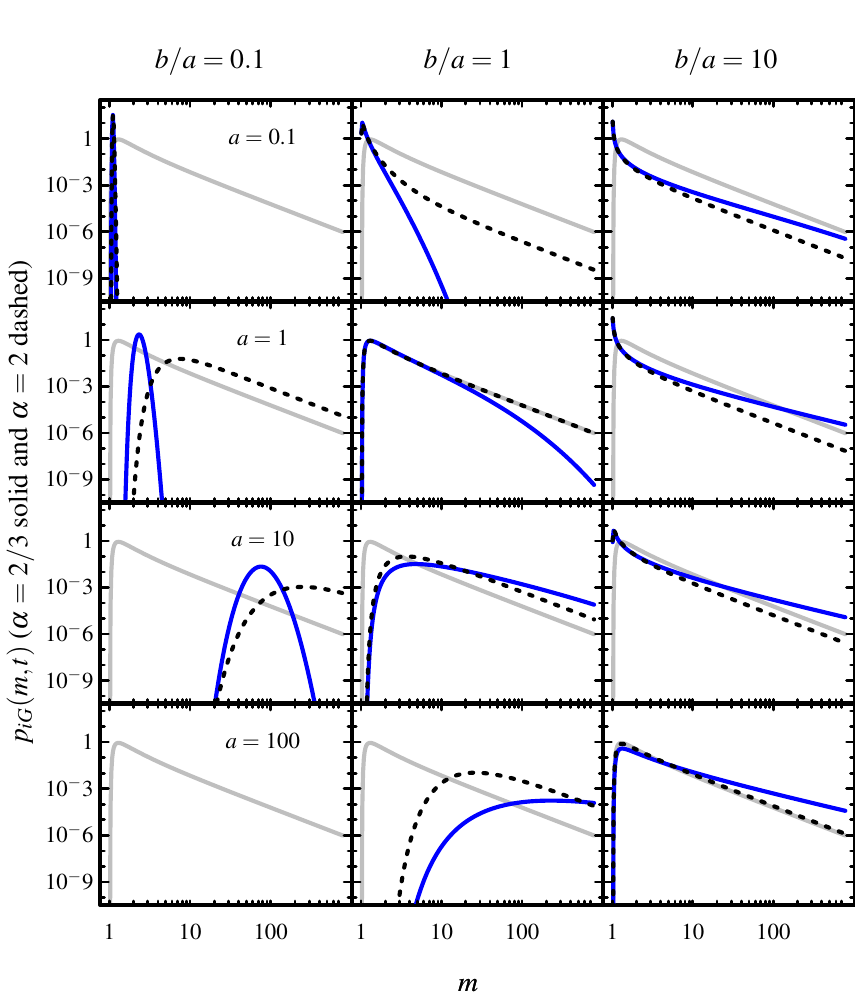}
\includegraphics[width=8.5cm]{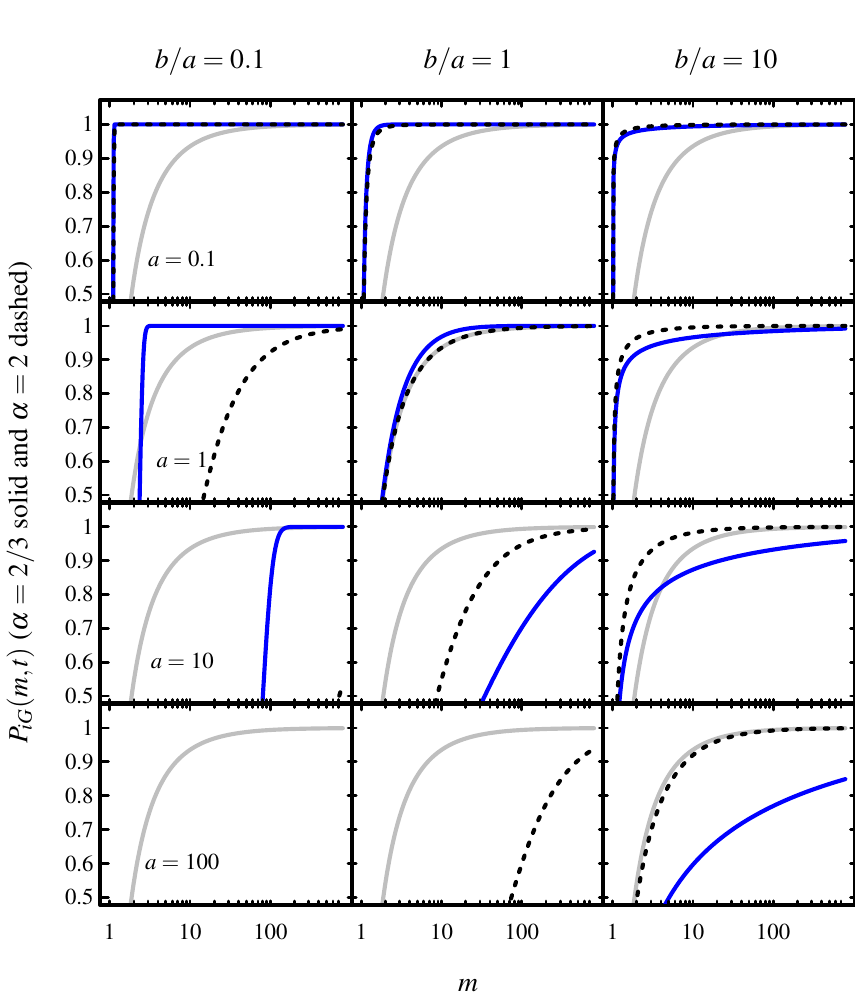}
\caption{ \label{figure_noise_comp}
Probability density $p_\text{iG}$ (left) and cumulative distribution $P_\text{iG}$ (right) for various $a$ and $b/a$ with $m_0=1$ and $t=1$.
The solid curve is for $\alpha = 2/3$ and the dotted curve for $\alpha=2$.
As reference each panel shows $p_\text{iG}$ and $P_\text{iG}$ for $\alpha=2$, $a=1$ and $b/a=1$ in grey.
}
\end{figure*}

\begin{table}
\caption{\label{table_scaling_factors} Some values for scaling of the initial mass and average accretion rate or time.} 
Scaling with initial mass:
$k_m=  (m_0'/m_0)^{1-\alpha}$, $ a' = k_m a$ and $b' = k_m b$ 
\setlength{\tabcolsep}{.8ex}
\begin{tabular}{llrrrrrrrrr}
 & $m_0'/m_0$ &  0.01 &  0.10 &  0.20 &  0.50 &  1.00 &  2.00 &  5.00 &  10.00  \\
\multicolumn{1}{l}{$\alpha=2$} & $k_m=$      & 100.00 &  10.00 &  5.00 &  2.00 &  1.00 &  0.50 &  0.20 &  0.10 \\
\multicolumn{1}{l}{$\alpha=2/3$} & $k_m=$       &  0.22 &  0.46 &  0.58 &  0.79 &  1.00 &  1.26 &  1.71 &  2.15
\end{tabular}
Scaling with average accretion rate
$ a' = k_a a$ , $b' = \sqrt{k_a} b$, $t' = t/k_a$
\begin{tabular}{lrrrrrrrrrr}
$a'=k_a a =$           &  0.10 &  0.20 &  0.50 &  1.00 &  2.00 &  5.00 &  10.00 &  \\
$t'=t/k_a$                &  10.00 &  5.00 &  2.00 &  1.00 &  0.50 &  0.20 &  0.10 &   \\
$b'=\sqrt{k_a} b=$  &  0.32 &  0.45 &  0.71 &  1.00 &  1.41 &  2.24 &  3.16 &   \\
$b'/a'=$               &  3.16 &  2.24 &  1.41 &  1.00 &  0.71 &  0.45 &  0.32 &    \\
\end{tabular}
\end{table}

Forming stars are embedded in the flocculent environment of their natal cloud and will accrete from it, but in contrast to Bondi-Hoyle accretion the ambient medium is self-gravitating as well.
\citet{BonnellClarkeBate-2001a} argue that in a gas-dominated potential the accretion rate should follow $m^{2/3}$, whereas in a stellar-dominated potential (uncorrelated velocities of gas and stars) classical Bondi-Hoyle accretion $\propto m^2$ occurs.
Fluctuations in the accretion rate of forming stars can be generated either by the turbulence of the gas, or by the self-gravity of the gas cloud which generates filaments, or by the chaotic orbital motion of the forming star, which will bring it into regions of different gas density.
Fluctuating accretion can determine the shape of the stellar initial mass function.
In this section, we investigate the range of values that parameters can take so that accretion has an impact.
A measurement of the fluctuations in numerical studies that account for the self-gravity of the gas is required to answer what are the ramifications of accretion in the star formation process.

Accretion can shape the initial mass function in two ways:
a power-law tail appears and the distribution becomes wider than the distribution of the initial masses.
 An additional distribution of growth times will contribute to both as well.
Accretion will only have an impact on mass functions if the initial masses are spread over a sufficiently wide range and populate a power-law tail.
This occurs if $p_\text{iG}$ has a rounded triangular shape.

In Fig. \ref{figure_noise_comp}, we show in the left plot the probability density, $p_\text{iG} (m, t=1)$, and in the right plot the cumulative distribution  for a range of average accretion rates $a$ and level of fluctuations $b/a$.
$\dot{m} \propto m^{2/3}$ is shown as solid curve and $\dot{m} \propto m^2$  as dotted curve.
Starting mass is $m_0 = 1$ and growth time is $t=1$.
$\alpha=2/3$ constrains both $a$ and $b/a$ to be of order unity, otherwise $p_\text{iG}$ develops a power-law part too shallow to be consistent with the initial mass function.
$\alpha=2$ allows for a larger range of parameters.

For both $\alpha$ the average accretion rate $a$ needs to be larger than unity in order to sufficiently spread out the initial masses.
Although a power-law tail appears and becomes stronger with increasing $b/a$, it does not contain many stars (cumulative distribution in the power-law tail already $\approx 1$).

For a mean accretion rate $a \gtrapprox 1$, the situation depends on the level of fluctuations.
If $b/a \ll 1$, then growth is effectively deterministic, which less affects growth with $\alpha=2$.
If the level of fluctuations is very large ($ b/a \gg 1 $), $p_\text{iG}$ is very peaked at $m_0$ and the power-law tail is not strongly populated as well, unless the average accretion rate is very high. 
For a lognormal distribution, the ratio between median and mean is $R= \exp ( - \sigma_l^2/2)=( (b/a)^2 + 1 )^{-1/2}$, depending only on the ratio of standard deviation to mean.
When $\sigma_l = 2$  ($b/a=7.32$), this means that the median accretion rate is only $\approx$ 1/10 of the mean accretion rate.
For comparison, $\sigma_l=1$ corresponds to $b/a = 1.31$.

With increasing $a$ and $b/a$ the exploding fraction has to be considered for $\alpha = 2$.
For $b/a=0.1$ it is not relevant at $t=1$ ($< 10^{-25}$), as for $a=0.1$. 
For $b/a =1$, the exploding fraction is $3 \times 10^{-5}$ for $a=1$, and 10 and 15\% for $a=10$ and $a=100$, respectively. 
For $b/a=10$ a significant fraction of the  seeds explode (34\% at $a=1$, 45\% at $a=10$ and $100$).
These parameter combinations would require some feedback mechanism to prohibit explosions.

Scaling to a different initial mass leaves the shape of $p_\text{iG} (m, t)$ unchanged for the same $t$, if $a$ and $b$ are scaled by multiplying both with the scaling factor $k_m=(m_0'/m_0)^{1-\alpha}$.
Values for the scaling factor for some typical initial masses are given in Table \ref{table_scaling_factors}.
For $\alpha=2$, they span a very wide range, whereas for $\alpha=2/3$ they lie around unity.

Scaling the mean accretion rate  does not preserve the shape of $p_\text{iG}$, unless the ratio $b/a$ and the time scale are changed.
Correspondingly, if the same shape of $p_\text{iG}$ should be achieved in a different time both $a$ and $b$ require scaling.
Here the scaling does not preserve the ratio $b/a$.
Some values for the scaled parameters are given in Table \ref{table_scaling_factors}.

The above findings can be summarized in the rule of thumb that both the average accretion rate $a$ and the level of fluctuations $b/a$ have to be of order unity for unit initial mass and unit growth time in order to sufficiently populate the power-law tail.
Nevertheless, there are other parameter combinations that may also lead to the desired behaviour of $p_\text{iG}$.
Furthermore, if the initial mass, the  average accretion rate and the time are given in physical units an observed ratio $b/a$ may be far from unity.

\subsection{Effects of a distribution of initial masses and growth times}\label{sec_imf}

\begin{figure}
\begin{center}
\includegraphics[width=6cm]{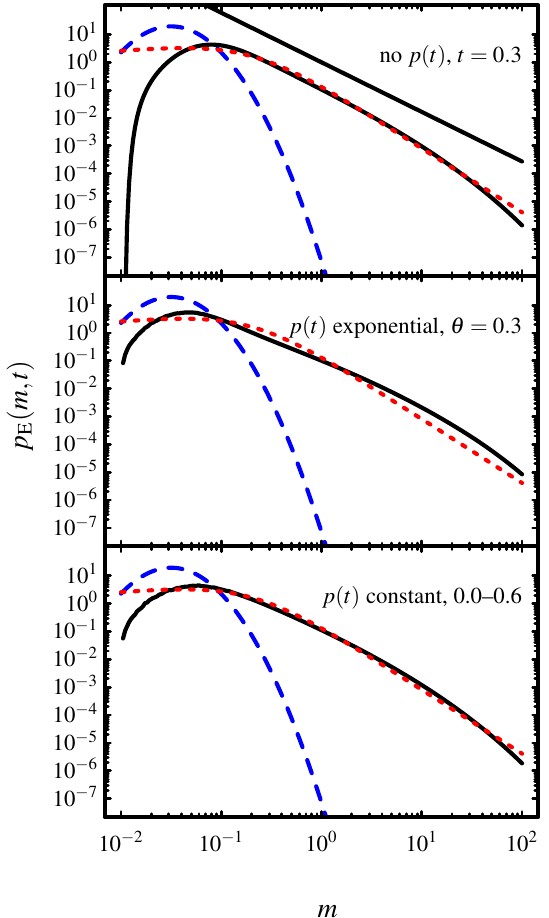}
\end{center}
\caption{ \label{figure_ensembles}
Ensemble distribution functions for a population growing from a distribution of initial masses (dashed) with different distribution of growth times.
$\alpha=2/3$, $a=2$ and $b=1$.
In the top panel, all stars accrete for the same time, the middle panel uses an exponential distribution of growth times and the bottom panel a uniform distribution.
The dotted line shows a standard system mass function.
}
\end{figure}

Fig. \ref{figure_ensembles} shows the effects of a distribution of initial masses and a distribution of growth times.
We use a lognormal distribution of initial masses (equation \ref{lognormal_pdf}) with $\mu_l = -3.15$ and $\sigma_l=0.55$, corresponding to a mean mass $\bar{m_0} = 0.05\ \Msun$.
$p(m_0)$ is shown as dashed curve in Fig. \ref{figure_ensembles}.
The distribution function for an ensemble growing from a distribution of initial masses is given by
\< p_\text{E} (m,t) &=& \int p(m_0) p_\text{iG} (m,t ; \alpha,a,b,m_0) \d m_0. \>
The top panel of Fig. \ref{figure_ensembles} shows $p_\text{E}$ for $\alpha=2/3$, $a=3\ \Msun^{1/3}/\text{Myr}$ and $b=\sqrt{3}\ \Msun^{1/3}/\text{Myr}$ at $t=0.3$ Myr, which corresponds to the dimensionless case of $a=1$, $b=1$ and $t=1$.
Note that here no explosions occur as $\alpha < 1$.
The peak of $p_\text{E}$ is significantly shifted compared to the distribution of the $m_0$.
Above $m_\beta=0.45\ \Msun$ (calculated using $\bar{m_0}$) a power-law part $\propto m^{-1.8}$ appears ($\beta= 1.11$) which starts to decay for larger masses.
At $m= 1\ \Msun$ the exponent is $-2.6$ and $-4.4$ at $m=10\ \Msun$.
For comparison we also show the system IMF as dotted (\citealp{Chabrier-2003b}; parametrization by\citealp{Maschberger-2013a}: $p_\text{Sys} \propto (m/\mu)^{-\alpha} \( 1 +(m/\mu)^{1-\alpha} \)^{-\beta}$, $\alpha=2.3$, $\beta=2$, $\mu=0.2$).
$p_\text{E}$ resembles $p_\text{Sys}$ for stellar masses on, but under-populates the brown dwarf region.

Not only the initial masses but also the time how long a star accretes can be distributed.
This may have an important impact on the ensemble distribution:
 \citet{BasuJones-2004a} found that an exponential distribution of growth times leads to a power-law tail instead of a lognormal distribution for linear growth with Gaussian fluctuations.
\citet{BateBonnell-2005a} have similar findings for constant growth with lognormal fluctuations.
The ensemble distribution function for both a distribution of initial masses and a distribution of growth times is given by
\< p_\text{E} (m) &=& \int \int p(m_0) p(t) p_\text{iG} (m,t; \alpha, a, b, m_0) \d m_0 \d t. \>
The middle panel of Fig. \ref{figure_ensembles} shows $p_\text{E}$ for a lognormal distribution of $m_0$, as above, and an exponential distribution of $t$ ($p(t) = \theta^{-1} \exp (- t/\theta)$).
We choose $\theta = 0.3$, which is the mean growth time.
At large masses the power law is flatter compared to the case without a distribution of growth times, populated by the stars that had more time to grow.
$p_\text{E}$ is visibly shallower than $p_\text{Sys}$.

A uniform distribution of growth times (bottom panel, $t$ between 0 and 0.6 Myr) does not have the long tail like an exponential distribution, so that the power-law part of $p_\text{E}$ is steeper as it is not populated by the stars growing for a very long time.
Here also the brown dwarf regime is more populated.

\begin{figure}
\begin{center}
\includegraphics[width=6cm]{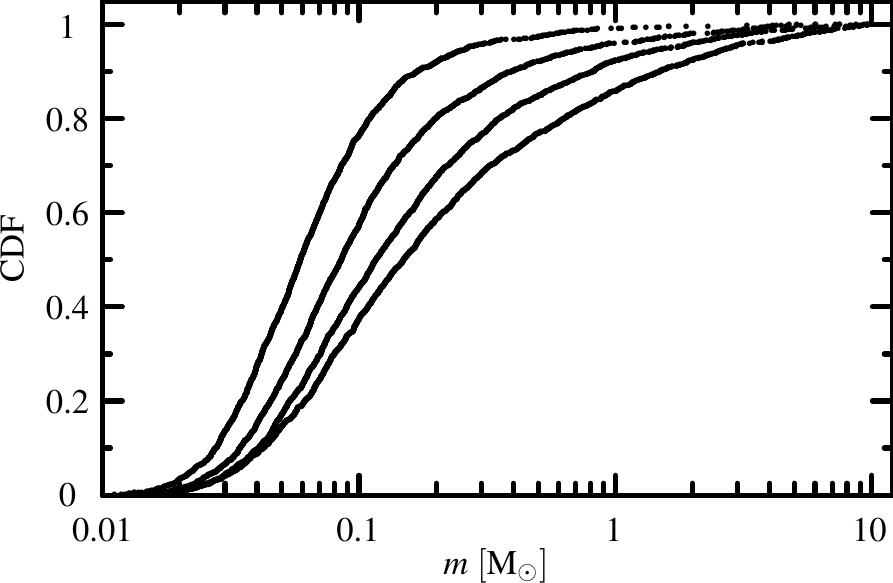}
\end{center}
\caption{ \label{figure_ensembles_time}
Cumulative ensemble distribution functions for a population growing with a distribution of initial masses and an exponential distribution of growth times (Corresponding to the middle panel of Fig. \ref{figure_ensembles}) stopping at $t=0.1$ Myr, $t=0.3$ Myr, $t=0.6$ Myr, and $t=1.0$ Myr (from left to right).
}
\end{figure}

With a distribution of initial masses and growth times the ensemble mass function is also evolving in time.
Figure \ref{figure_ensembles_time} shows the cumulative distribution of $p_\text{E}$ with an exponential distribution of growth times stopping at $t=0.1$, $0.3$, $0.6$ and $1.0$ Myr (left to right curves).
Rejection sampling has been used to obtain a sample $p_\text{E}$ containing 2000 variates.
The mean masses are  $\bar{m}=0.11$ ,$0.24$, $0.39$, and $0.61\ \Msun$, and their time evolution not dissimilar to the evolution of the mean mass in hydrodynamical simulations of star formation
\citep{BonnellVineBate-2004a,MaschbergerClarkeBonnell-2010a,Bate-2012a,KrumholzKleinMcKee-2012a}
The cumulative distribution functions seem to evolve somewhat more in time than in simulations, \citep{KrumholzKleinMcKee-2012a,Bate-2012a}.

The parameters used in this section are chosen for illustrating purposes such that the resulting $p_\text{E}$ has some resemblance of the observed system mass function.
Although the parameters have some reasonable values (for example, a star grows for a few hundred thousand years), it is very necessary to infer them from numerical simulations of star formation in order to constrain the scenario.

\section{Summary}

We investigate the consequences of fluctuations in the accretion rates of non-linearly accreting stars  by the means of a non-linear multiplicative stochastic process and find the following:

\begin{enumerate}

\item Non-linear accretion, $\dot{m} \propto m^\alpha$, with fluctuations in the accretion rates lead to power-law tails in the distribution function of the final masses.

\item The main body of the distribution of final masses resembles a lognormal distribution.

\item Gaussian fluctuations produce a power law tail $\propto m^{-\alpha}$  at high masses for $\alpha > 1$ and at low masses for $\alpha < 1$.

\item Lognormal fluctuations, approximated by the inverse Gaussian distribution, always produce a mass distribution function that has a power-law tail at high masses, even if accretion is linear ($\dot{m} \propto m$) or the $\alpha < 1$. Only for $\alpha \ge 1$, the exponent of the power-law tail is not evolving in time. For $\alpha < 1$, the power-law tail decays at very large masses.

\item The shape of the mass distribution function depends on the initial mass, the average accretion rate $a$, the amount of fluctuations (ratio between standard deviation and average accretion rate $b/a$) and time.

\item The power-law tail is more and more populated in time, similar to time-evolution of the mass function obtained in numerical simulations of star formation. If observed at an early time the power-law tail may appear steeper because it is only sparsely sampled.

\item The final distribution function can resemble the whole shape of the initial mass function.

\end{enumerate}

\section{Acknowledgements}
I acknowledge funding via the ANR 2010 JCJC 0501 1 ``DESC'' (Dynamical Evolution of Stellar Clusters).
I would like to thank Cathie Clarke for very insightful discussions, as well as Estelle Moraux and Christophe Becker for their fruitful comments.

\bibliographystyle{mn2e}
\bibliography{refs_tm}

\begin{thebibliography}{}

\bibitem[\protect\citeauthoryear{{Basu} \& {Jones}}{{Basu} \&
  {Jones}}{2004}]{BasuJones-2004a}
{Basu} S.,  {Jones} C.~E.,  2004, \mnras, 347, L47

\bibitem[\protect\citeauthoryear{{Basu} \& {Jones}}{{Basu} \&
  {Jones}}{2005}]{BasuJones-2005a}
{Basu} S.,  {Jones} C.~E.,  2005, in {E.~Corbelli, F.~Palla, \& H.~Zinnecker}
  ed., The Initial Mass Function 50 Years Later Vol.~327 of Astrophysics and
  Space Science Library, {A minimum hypothesis explanation for an IMF with a
  lognormal body and power law tail}.
pp 437--+

\bibitem[\protect\citeauthoryear{{Bate}}{{Bate}}{2012}]{Bate-2012a}
{Bate} M.~R.,  2012, \mnras, 419, 3115

\bibitem[\protect\citeauthoryear{{Bate} \& {Bonnell}}{{Bate} \&
  {Bonnell}}{2005}]{BateBonnell-2005a}
{Bate} M.~R.,  {Bonnell} I.~A.,  2005, \mnras, 356, 1201

\bibitem[\protect\citeauthoryear{{Bate}, {Bonnell} \& {Bromm}}{{Bate}
  et~al.}{2003}]{BateBonnellBromm-2003a}
{Bate} M.~R.,  {Bonnell} I.~A.,    {Bromm} V.,  2003, \mnras, 339, 577

\bibitem[\protect\citeauthoryear{{Bonnell} \& {Bate}}{{Bonnell} \&
  {Bate}}{2006}]{BonnellBate-2006b}
{Bonnell} I.~A.,  {Bate} M.~R.,  2006, \mnras, 370, 488

\bibitem[\protect\citeauthoryear{{Bonnell}, {Bate}, {Clarke} \&
  {Pringle}}{{Bonnell} et~al.}{1997}]{BonnellBateClarke-1997a}
{Bonnell} I.~A.,  {Bate} M.~R.,  {Clarke} C.~J.,    {Pringle} J.~E.,  1997,
  \mnras, 285, 201

\bibitem[\protect\citeauthoryear{{Bonnell}, {Bate}, {Clarke} \&
  {Pringle}}{{Bonnell} et~al.}{2001a}]{BonnellBateClarke-2001a}
{Bonnell} I.~A.,  {Bate} M.~R.,  {Clarke} C.~J.,    {Pringle} J.~E.,  2001,
  \mnras, 323, 785

\bibitem[\protect\citeauthoryear{{Bonnell}, {Clarke}, {Bate} \&
  {Pringle}}{{Bonnell} et~al.}{2001b}]{BonnellClarkeBate-2001a}
{Bonnell} I.~A.,  {Clarke} C.~J.,  {Bate} M.~R.,    {Pringle} J.~E.,  2001,
  \mnras, 324, 573

\bibitem[\protect\citeauthoryear{{Bonnell}, {Vine} \& {Bate}}{{Bonnell}
  et~al.}{2004}]{BonnellVineBate-2004a}
{Bonnell} I.~A.,  {Vine} S.~G.,    {Bate} M.~R.,  2004, \mnras, 349, 735

\bibitem[\protect\citeauthoryear{{Cartwright} \& {Whitworth}}{{Cartwright} \&
  {Whitworth}}{2012}]{CartwrightWhitworth-2012a}
{Cartwright} A.,  {Whitworth} A.~P.,  2012, \mnras, 423, 1018

\bibitem[\protect\citeauthoryear{{Chabrier}}{{Chabrier}}{2003}]{Chabrier-2003b}
{Chabrier} G.,  2003, \pasp, 115, 763

\bibitem[\protect\citeauthoryear{Chandrasekhar}{Chandrasekhar}{1943}]{Chandrasekhar-1943a}
Chandrasekhar S.,  1943, Reviews of Modern Physics, 15, 1

\bibitem[\protect\citeauthoryear{{Dib}, {Kim} \& {Shadmehri}}{{Dib}
  et~al.}{2007}]{DibKimShadmehri-2007a}
{Dib} S.,  {Kim} J.,    {Shadmehri} M.,  2007, \mnras, 381, L40

\bibitem[\protect\citeauthoryear{{Dib}, {Shadmehri}, {Padoan}, {Maheswar},
  {Ojha} \& {Khajenabi}}{{Dib} et~al.}{2010}]{DibShadmehriPadoan-2010a}
{Dib} S.,  {Shadmehri} M.,  {Padoan} P.,  {Maheswar} G.,  {Ojha} D.~K.,
  {Khajenabi} F.,  2010, \mnras, 405, 401

\bibitem[\protect\citeauthoryear{{Edgar}}{{Edgar}}{2004}]{Edgar-2004a}
{Edgar} R.,  2004, New Astronomy Reviews, 48, 843

\bibitem[\protect\citeauthoryear{{Elmegreen} \& {Mathieu}}{{Elmegreen} \&
  {Mathieu}}{1983}]{ElmegreenMathieu-1983a}
{Elmegreen} B.~G.,  {Mathieu} R.~D.,  1983, \mnras, 203, 305

\bibitem[\protect\citeauthoryear{Filippov}{Filippov}{1961}]{Filippov-1961b}
Filippov A.~F.,  1961, Theory Of Probability And Its Applications, 6, 275

\bibitem[\protect\citeauthoryear{{Klessen} \& {Burkert}}{{Klessen} \&
  {Burkert}}{2000}]{KlessenBurkert-2000a}
{Klessen} R.~S.,  {Burkert} A.,  2000, \apjs, 128, 287

\bibitem[\protect\citeauthoryear{{Kolmogorov}}{{Kolmogorov}}{1941}]{Kolmogorov-1941a}
{Kolmogorov} A.~N.,  1941, Dokl. Akad. Nauk SSSR, 31, 99

\bibitem[\protect\citeauthoryear{{Kolmogorov}}{{Kolmogorov}}{1992}]{Kolmogorov-1992a}
{Kolmogorov} A.~N.,  1992, in Shiryayev A.~N.,  ed., Selected Works of A. N.
  Kolmogorov Vol. 2 Probability Theory and Mathematical Statistics, {On the
  logarithmic normal distribution of particle sizes under grinding}.
Kluwer, pp 282--284

\bibitem[\protect\citeauthoryear{{Kroupa}}{{Kroupa}}{2002}]{Kroupa-2002a}
{Kroupa} P.,  2002, Science, 295, 82

\bibitem[\protect\citeauthoryear{{Krumholz}, {Klein} \& {McKee}}{{Krumholz}
  et~al.}{2012}]{KrumholzKleinMcKee-2012a}
{Krumholz} M.~R.,  {Klein} R.~I.,    {McKee} C.~F.,  2012, \apj, 754, 71

\bibitem[\protect\citeauthoryear{Krumholz, McKee \& Klein}{Krumholz
  et~al.}{2005}]{KrumholzMcKeeKlein-2005b}
Krumholz M.~R.,  McKee C.~F.,    Klein R.~I.,  2005, Nature, 438, 332

\bibitem[\protect\citeauthoryear{{Krumholz}, {McKee} \& {Klein}}{{Krumholz}
  et~al.}{2006}]{KrumholzMcKeeKlein-2006a}
{Krumholz} M.~R.,  {McKee} C.~F.,    {Klein} R.~I.,  2006, \apj, 638, 369

\bibitem[\protect\citeauthoryear{{Larson}}{{Larson}}{1973}]{Larson-1973a}
{Larson} R.~B.,  1973, \mnras, 161, 133

\bibitem[\protect\citeauthoryear{{Larson}}{{Larson}}{1978}]{Larson-1978a}
{Larson} R.~B.,  1978, \mnras, 184, 69

\bibitem[\protect\citeauthoryear{{Marcus}}{{Marcus}}{1968}]{Marcus-1968a}
{Marcus} A.~H.,  1968, Technical report, {Random independent splitting model
  for the mass spectrum of protostars and interstellar clouds}.
Bellcomm, Inc

\bibitem[\protect\citeauthoryear{{Maschberger}}{{Maschberger}}{2013}]{Maschberger-2013a}
{Maschberger} T.,  2013, \mnras, 429, 1725

\bibitem[\protect\citeauthoryear{{Maschberger}, {Clarke}, {Bonnell} \&
  {Kroupa}}{{Maschberger} et~al.}{2010}]{MaschbergerClarkeBonnell-2010a}
{Maschberger} T.,  {Clarke} C.~J.,  {Bonnell} I.~A.,    {Kroupa} P.,  2010,
  \mnras, 404, 1061

\bibitem[\protect\citeauthoryear{{Myers}}{{Myers}}{2000}]{Myers-2000a}
{Myers} P.~C.,  2000, \apjl, 530, L119

\bibitem[\protect\citeauthoryear{{Myers}}{{Myers}}{2008}]{Myers-2008a}
{Myers} P.~C.,  2008, \apj, 687, 340

\bibitem[\protect\citeauthoryear{{Myers}}{{Myers}}{2009}]{Myers-2009a}
{Myers} P.~C.,  2009, \apj, 706, 1341

\bibitem[\protect\citeauthoryear{{Myers}}{{Myers}}{2011}]{Myers-2011a}
{Myers} P.~C.,  2011, \apj, 743, 98

\bibitem[\protect\citeauthoryear{{Myers}}{{Myers}}{2012}]{Myers-2012a}
{Myers} P.~C.,  2012, \apj, 752, 9

\bibitem[\protect\citeauthoryear{{Nordlund} \& {Padoan}}{{Nordlund} \&
  {Padoan}}{1999}]{NordlundPadoan-1999a}
{Nordlund} {\AA}.~K.,  {Padoan} P.,  1999, in {Franco} J.,  {Carraminana} A.,
  eds, Interstellar Turbulence {The Density PDFs of Supersonic Random Flows}.
p.~218

\bibitem[\protect\citeauthoryear{{\O}ksendal}{{\O}ksendal}{2002}]{Oksendal-2002a}
{\O}ksendal B.,  2002, Stochastic Differential Equations.
Springer-Verlag

\bibitem[\protect\citeauthoryear{{Passot} \& {V{\'a}zquez-Semadeni}}{{Passot}
  \& {V{\'a}zquez-Semadeni}}{1998}]{PassotVazquez-Semadeni-1998a}
{Passot} T.,  {V{\'a}zquez-Semadeni} E.,  1998, \pre, 58, 4501

\bibitem[\protect\citeauthoryear{Thorin}{Thorin}{1977}]{Thorin-1977a}
Thorin O.,  1977, Scandinavian Actuarial Journal, 1977, 121

\bibitem[\protect\citeauthoryear{van Kampen}{van Kampen}{2006}]{Kampen-2006a}
van Kampen N.,  2006, Stochastic Processes in Physics and Chemistry.
Elsevier, Amsterdam

\bibitem[\protect\citeauthoryear{{Vazquez-Semadeni}}{{Vazquez-Semadeni}}{1994}]{Vazquez-Semadeni-1994a}
{Vazquez-Semadeni} E.,  1994, \apj, 423, 681

\bibitem[\protect\citeauthoryear{{Zinnecker}}{{Zinnecker}}{1982}]{Zinnecker-1982a}
{Zinnecker} H.,  1982, Annals of the New York Academy of Sciences, 395, 226

\bibitem[\protect\citeauthoryear{{Zinnecker}}{{Zinnecker}}{1984}]{Zinnecker-1984a}
{Zinnecker} H.,  1984, \mnras, 210, 43

\end{thebibliography}
\label{lastpage}
\bsp
\end{document}